\documentclass{aastex7}
\usepackage{amsmath}
\usepackage{graphics}
\usepackage{booktabs}
\usepackage{siunitx}

\begin{document}

\title{Cosmic Ray Detection and Rejection for CSST}

\author[orcid=0000-0002-7210-8104]{Yan Yu}
\affiliation{School of Physics and Astronomy, Sun Yat-sen University, Zhuhai, China}
\affiliation{National Astronomical Observatories, Chinese Academy of Sciences, Beijing, China}
\email{yuyan35@mail2.sysu.edu.cn}  

\author[0000-0002-6077-6287]{Bin Ma}
\affiliation{School of Physics and Astronomy, Sun Yat-sen University, Zhuhai, China}
\affiliation{CSST Science Center for the Guangdong-Hong Kong-Macau Greater Bay Area, Zhuhai 519082, China}
\email{mabin3@mail.sysu.edu.cn}

\author[0000-0002-8531-5161]{Tianmeng Zhang}
\affiliation{School of Astronomy and Space Science, University of Chinese Academy of Sciences, Beijing 101408, People’s Republic of China}
\affiliation{Key Laboratory of Space Astronomy and Technology, National Astronomical Observatories, Chinese Academy of Sciences, 20A Datun Road, Beijing 100101, People’s Republic of China}
\email{zhangtm@nao.cas.cn}

\author{Yi Hu}
\affiliation{National Astronomical Observatories, Chinese Academy of Sciences, Beijing, China}
\email{huyi.naoc@gmail.com}

\author{Yajie Zhang}
\affiliation{College of Intelligence and Computing, Tianjin University, Tianjin 300350, China }
\affiliation{Technical R\&D Innovation Center, National Astronomical Data Center, Tianjin 300350, China}
\email{zyj0928@tju.edu.cn}

\correspondingauthor{Bin Ma}
\email{mabin3@mail.sysu.edu.cn}

\correspondingauthor{Tianmeng Zhang}
\email{zhangtm@nao.cas.cn}


\begin{abstract}
As a space telescope, the China Space Station Survey Telescope (CSST) will face significant challenges from cosmic ray (CR) contamination. These CRs will severely degrade image quality and further influence scientific analysis. Due to the CSST's sky survey strategy, traditional multi-frame stacking methods become invalid. The limited revisits prompted us to develop an effective single-image CR processing method for CSST. We retrained the DeepCR model based on CSST simulated images and achieved $97.90 \pm 0.18\%$ recall and 98.67 ± 0.05\% precision on CR detection.
Moreover, this paper puts forward an innovative morphology-sensitive inpainting method, which focuses more on areas with higher scientific value.  We trained a UNet++ model especially on contaminated stellar/galactic areas, alongside adaptive median filtering for background regions. This method achieves effective for CRs with different intensities and  different distances from centers of scientific targets. By this approach, the photometric errors of CR-corrected targets could be restricted to the level comparable to those of uncontaminated sources. Also, it increases the detection rate by 13.6\% compared to CR masking. This method will provide a robust CR mitigation for next-generation space telescopes.
\end{abstract}


\section{Introduction}
As an important next-generation astronomical observation facility, the China Space Station Survey Telescope (CSST) is China's first large-scale optical astronomical telescope located in outer space \citep{RN60}. Operating in the unique space environment, this space-based telescope faces distinct challenges compared to ground-based ones. Without atmospheric protection, CSST will be exposed to a large amount of cosmic rays(CRs), which can severely contaminate its images. CR contamination occurs when charged particles in CRs pass through solid-state detectors, such as CCD or CMOS sensors. These particles generate electron-hole pairs, causing accumulation of excess charges in the affected pixels. CRs from different sources leave signals with various intensities on detectors. In astronomical images, CRs appear as randomly distributed over-bright pixels with distinct and sharp edges. Depending on the incident angle, CRs typically manifest as point-like features that may span several pixels, as well as elongated track-like or worm-like patterns that can extend from tens to nearly a hundred pixels \citep{RN11}.

CR contamination introduces significant noise in astronomical observations, which greatly degrades image quality and compromises subsequent scientific analyses \citep{RN10, RN48}. According to \citet{RN51}, a single CR event affects approximately 7 pixels on average and deposits about 2700 electrons on the Hubble Space Telescope (HST). These values are specific to the HST's detectors and can vary significantly depending on the size and thickness of the image sensor used. This study also provides complete statistics on the total events, their morphologies, and the lengths of contamination paths across various instruments. Furthermore, the Euclid VIS instrument (with 566~s exposures) Quick Data Release (Q1) shows an average CR contamination rate of 2.05\%. In some localized areas, this rate reaches as high as 5\%, displaying characteristic "X-ray" spatial patterns \citep{collaboration2025euclid}. As for the CSST, with 150 seconds exposures, simulation results indicate that approximately 200,000 pixels per image are affected by CRs, which accounts for 2–3\% of the effective pixels in the astronomical observations (Fang et al., 2025, submitted). Therefore, effective mitigation of these artifacts is critical to maintaining the telescope's image quality and the validity of research findings.

Utilizing the fact that CRs rarely strike the same pixel in multiple exposures, CR removal can be easily achieved through multi-frame image stacking analysis. By comparing several (at least 3) distinct frames, contaminated pixels can be identified and replaced using median values from unaffected frames, such as \textit{AstroDrizzle} did for the HST images \citep{RN28}. However, the CSST wide-field survey strategy provides only a single pass per detector for any given sky region. Although observations in each band (utilizing 2-4 independent detectors) yield 2-4 revisits, these exposures exhibit significant temporal discontinuities. Such sparse and non-contiguous observations make it impossible to remove CRs using multi-frame median stacking. Besides, multi-frame stacking methods face serious limitations: they become improper when observations lack consecutive exposures or when the target's variability timescale is comparable to the exposure time. Additionally, variations in the point spread function (PSF) between exposures can lead to systematic errors in traditional multi-frame methods \citep{RN29}. Consequently, CR correction methods that can operate on individual images must be developed and deployed.

Traditional methods of single-frame CR detection could be primarily categorized into two classes: statistical and morphological. Statistical approaches include sigma-thresholding for outlier identification based on the standard deviation of pixel values. And the histogram-based method \citep{RN19} locates outliers based on assumptions of a Gaussian distribution. \citet{RN29} developed PSF-matched filtering for CR detection, because CRs are not affected by the telescope's PSF. Morphological methods, on the other hand, include the L.A.Cosmic algorithm \citep{RN15}, which detects CRs by identifying sharp-edged features through Laplace edge detection and analyzing local symmetry patterns, effectively detecting CRs from under-sampled point sources. Alternatively, the fuzzy logic approach \citep{RN32} describes images using three key characteristics (target size, edge gradient sharpness, and overall image sharpness), and detects CRs by simulating a human's visual inspection. 

Recent advances in machine learning have significantly improved CR detection accuracy and adaptability. Early studies used decision trees \citep{RN30} and artificial neural networks \citep{RN34} for star-CR classification. Deep learning breakthroughs include UNet based \textit{DeepCR} model \citep{RN20}, which enables end-to-end detection and inpainting, and on this basis \textit{Cosmic-CoNN} network \citep{RN38} addresses instrument dependency and class imbalance issues through enhanced generalization capabilities. \citet{2024ApJ...962....7C} also improved the DeepCR model by utilizing CRs in dark frames to eliminate the need for multiple frames of the same fields. Their retrained DeepCR model can detect approximately 7\% more good-quality stars. These developments provide optimized solutions for CR processing in next-generation telescopes such as CSST. While current CR detection methods demonstrate reliable performance through diverse algorithmic approaches, CR inpainting remains predominantly reliant on simple median/mean filtering techniques, with DeepCR's machine learning approach being a notable exception. These conventional methods exhibit critical limitations in repairing large-area CR contamination, and particularly, when CRs overlap scientifically vital regions such as galaxies or stars—precisely the primary targets of astronomical investigations rather than full-frame background areas. 

Several general image restoration techniques can also be applied to CR removal. For instance, OpenCV's built-in inpaint function offers two modes based on the Fast Marching Method (FMM) and Navier-Stokes equations, suitable for repairing small-area discrete CR hits.  Additionally, \textit{maskfill} \citep{RN14} provides a robust method for filling masked regions in astronomical images. Machine learning approaches include a generative adversarial network (GAN) model \citep{RN24} that achieves simultaneous noise suppression and structural recovery via adversarial training, and the \textit{Pix2WGAN} framework \citep{RN43} which improves image resolution while denoising. Probabilistic methods encompass the \textit{NIFTY} package \citep{RN41}, utilizing Gaussian processes and variational inference for image processing with superior performance on Gaussian noise profiles, and \textit{Astrofix} \citep{RN16}, which implements PSF-informed Gaussian process filtering for optical astronomy, achieving high accuracy in galaxy morphology recovery.

While various image restoration techniques have broad applicability across different fields, their effectiveness in correcting CR issues—particularly inpainting within stars or galaxies—remains limited.  Currently, there are no effective solutions for CR inpainting specifically aimed at scientifically important areas. Thus, we developed a specific inpainting method to inpaint contaminated stellar or galactic areas. This approach allows for reliable CR correction and ensures photometric precision of stars and galaxies, which are needed for astronomical analysis.

This paper is structured as follows: we first present the training datasets for CR processing in Section 2. Section 3 details our methodology for CR identification and correction. We then report comprehensive results in Section 4, including the accuracy of CR identification, correction efficacy on individual galaxies, full-frame performance on CSST images, and impacts on star/galaxy number count distributions. Finally, Section 5 summarizes the key findings, discusses limitations of the current approach, and outlines potential directions for future work.

\section{Dataset}

The data used in this article are created by CSST simulations pipeline. The detailed methodology for generating these simulated data is described in \citet{CSSTMOCK} (submitted). For the cosmic ray simulation, in brief, an initial CR impact fraction of $\sim$0.75‰ was adopted, with event lengths distributed between 0--100 pixels, event energy following a log-Gaussian distribution (mean $\log_{10}E = 3.3$, FWHM = 0.6), and charge diffusion modeled via Gaussian convolution ($\sigma$ = 0.2 pixels).

The CSST data processing pipeline consists of three stages: Level 0 to Level 2. In the CSST simulation, Level 0 data include the raw images and other input data, such as the input star catalogues and the simulated CRs data. Level 1 data are processed from Level 0. They have undergone instrumental effect corrections and various calibrations and are normalized to the unit of electrons per second (e$^{-}$/s). Additionally, they include FLAG images that identify bad pixels and CRs. Level 2 data involves catalogues, which are calculated based on the images after CR correction.

One of the major characteristics of CSST is that it can conduct multi-band photometry and slitless spectroscopic observations simultaneously. Its focal plane contains seven imaging bands ($NUV$, $u$, $g$, $r$, $i$, $z$, $y$) and three slitless spectroscopic bands ($GU$, $GV$, $GI$), with each band corresponding to two or more detectors, totaling 30 detectors (18 multi-band imaging detectors and 12 slitless spectroscopic detectors). Each exposure acquires data from these 30 detectors simultaneously \citep{RN60}. This study focuses exclusively on the 18 multi-band imaging detectors.

The dataset used in this paper consists of simulated CSST images, where Level 1 science data serve as base images, accompanied by CR masks derived from Level 0 raw data. To validate photometric accuracy during the CR correction analysis, the Level 0 input catalogues are also utilized.

All current training relies exclusively on simulated data. Once the telescope is operational in orbit, we will acquire multiple continuous exposures over several fixed sky regions to get true CRs by stacking those images. The utilization of CRs present in the calibrated dark frames provides another promising approach \citep{2024ApJ...962....7C}, which does not take up observation time. These real observational data will then be used to retrain our CR processing models. Consequently, the present study utilizes a limited number of simulated images, prioritizing diversity in sky coverage over quantity.

\subsection{The Development Dataset for CR Detection}
 
During the training of the CR identification model, a total of 288 image sets were utilised (16 sets per detector). Each set comprises a Level 1 science image and its corresponding Level 0 CRs data. For all 18 multi-band photometry detectors, we systematically constructed training sets and test sets for the DeepCR model. The dataset includes simulated images from eight representative fields with different number densities. Two simulated images per field were randomly selected to ensure comprehensive coverage of observational scenarios during model training. 

During training, only CRs with an intensity greater than 1 $e^-$ are selected to generate mask images. The detailed reasons for excluding 1 $e^-$ CRs will be introduced in Section \ref{sec:detection-result}. The current threshold of 1 $e^-$ can be adjusted in future training sessions to achieve the best training results.

\subsection{The Development Dataset for CR Rejection}\label{sec:inpaint_dataset}

This study focuses on star- and galaxy-centric CR correction, necessitating specialised training data. Since CRs rarely strike at astrophysical source centres, which results in samples being sparse and statistically biased. To address this, we generate a customised dataset: we extract 64×64-pixel sub-images centred on the centroids of galaxies or stars from full scientific images, and then artificially inject diverse CRs into varying locations within these sub-images. This 64×64-pixel size balances the spatial extent of typical stars/galaxies with computational efficiency, and can encapsulate most stellar profiles while capturing the core regions of extended galaxies.

\begin{figure*}[ht!]
\plotone{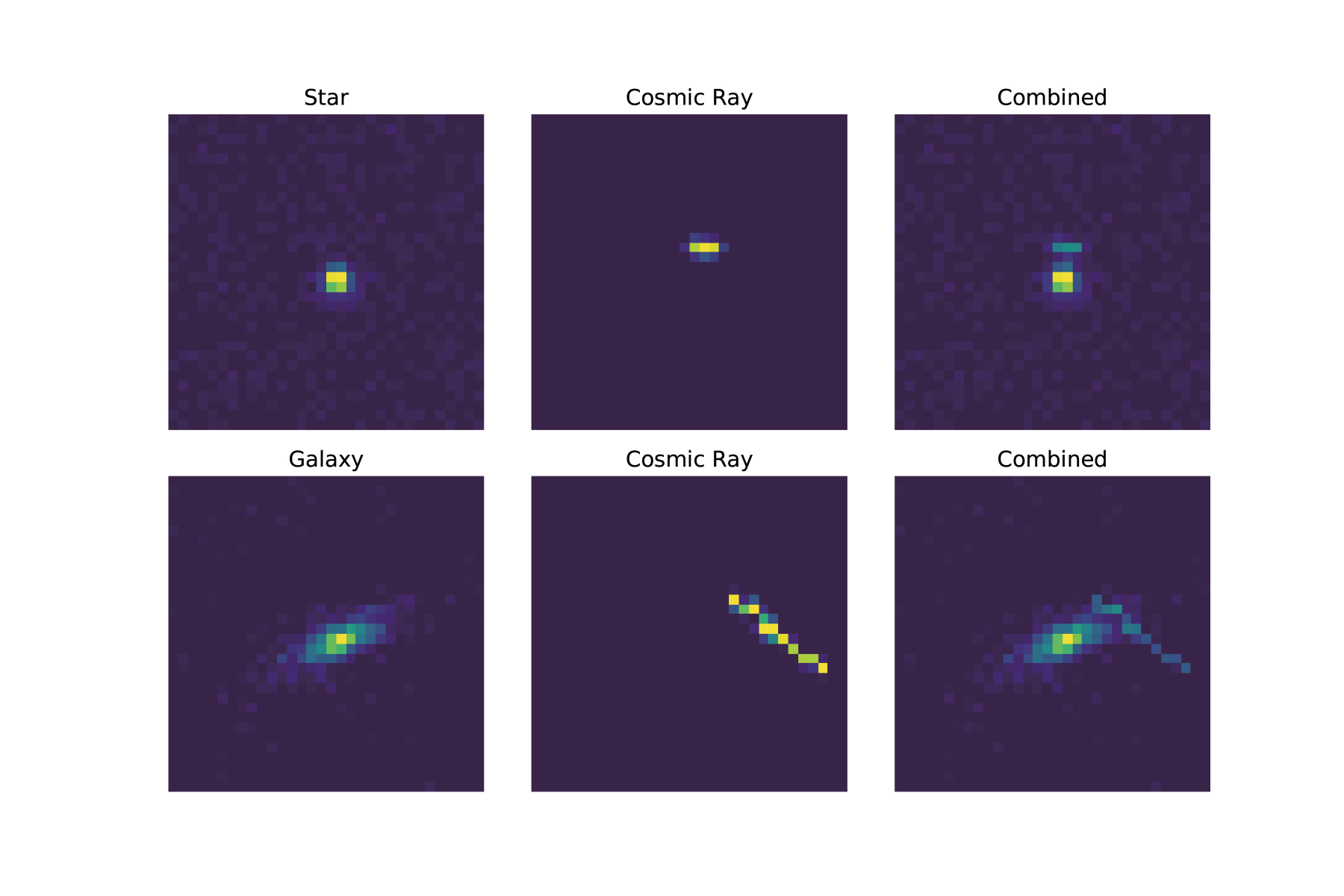}
\caption{The examples of the training dataset for CR rejection. The top row displays a star while the bottom row shows one galaxy. The left column features true images without CRs, the middle column shows CR samples (after shifting), and the right column displays combined images with CR interference.}
\label{fig:inpaintset}
\end{figure*}

We randomly selected 72 sets of full-frame images (4 sets per detector) spanning all 18 multi-band imaging detectors, each set comprising a Level 1 science image (including flags) and its corresponding Level 0 CR data. During image reductions from Level 0 to Level 1, essential calibrations, including flat-fielding and dark current subtraction, introduce a nonlinear mapping between pixels. This nonlinearity prevents us from using CR flux values from Level 0 data to reflect CR's intensity. Consequently, while CR masks are generated using Level 0 truth data, the actual CR-contaminated pixel values are sampled from the Level 1 science images. The dataset construction workflow involves the following:

\begin{enumerate}
\item Binarize CR signal to create full-frame CR masks.
\item Identify astrophysical sources from background-subtracted images, selecting source samples (column 1, Figure \ref{fig:inpaintset}) that require: (i) spatial extent \textgreater 10 pixels (excluding faint, unrecoverable sources) and (ii) CR-free 64×64 regions around source centers. Approximately 150 samples are ultimately selected from each image.
\item Extract isolated CRs (non-overlapping with sources) from background-subtracted images and crop 64×64 regions around their geometric centers as the CR samples.
\item Inject cosmic rays into the source samples. For each source sample, randomly select one CR from the CR samples (belonging to the same detector), then move the CR randomly by 0-10 pixels in both the x and y directions (column 2, Figure \ref{fig:inpaintset}), and overlay it onto the source image(column 3, Figure \ref{fig:inpaintset}).
\item Normalize the data using Min-Max scaling. This ensures consistent feature scales in the images while preserving the relative relationships of the data, thereby enhancing the convergence speed and stability of the machine learning model.
\end{enumerate}

\subsection{Scientific Test Set for CR Rejection}\label{sec:testset}

We developed an additional test dataset to evaluate photometric performance after CR rejection. The method is similar to that in Sect. \ref{sec:inpaint_dataset}. For each detector, we selected four sources whose magnitudes are approximately 17, 19, 21, and 23, respectively. And we chose 100 CR image clips per detector, covering a range of intensities. We then combined the selected sources and CRs by injecting 100 CR events with different intensities into each source image. These CR events were placed at radial distances between 0 and 15 pixels from the source centers. Finally, we saved the resulting synthetic images as FITS files. The FITS headers contain essential information for analysis, including the injected CR intensity level, the source-to-CR distance, the source magnitude, the photometric zeropoint, and the background noise levels.

\section{Method}

\subsection{CR Detection}
Through comparative analysis of CR identification methods, we selected the DeepCR deep learning model \citep{RN20}. This enhanced U-Net architecture employs encoder-decoder structures with skip connections to achieve pixel-level CR detection, demonstrating \textgreater 95\% accuracy in HST datasets.

We trained dedicated CR detection models for each detector of multi-band imaging following the DeepCR training protocol \footnote{https://deepcr.readthedocs.io/en/latest/index.html}, addressing distinct noise characteristics across bands. All models can be directly invoked via the CR detection functions provided by the DeepCR's standard API. Upon CR identification, detected CR pixels are assigned a dedicated value ($2^{20}$) in FLAG image, which is the third extension (ext3) of the Level 1 FITS files.

\subsection{Inpainting Method}

Although the deepCR model includes both mask and inpaint components, the inpaint component only provides a model for HST ACS-WFC-F606W, which is not suitable for images. Additionally, the specific training code is not publicly available. Therefore, we did not use deepCR-inpaint.

The median filtering algorithm can already achieve good results for CRs located in the background. It has strong robustness and a fast processing speed. However, for CRs that fall on the astrophysical sources, the median filter has obvious shortcomings. Thus, this study implements a tiered processing framework to balance precision and efficiency. 

This study employs the UNet++ model \citep{RN44} to repair CRs falling on the astronomical sources. The UNet++ architecture enhances the classic U-shaped encoder-decoder structure of UNet through dense nested skip pathways and deep supervision, improving feature fusion across network levels. These refined connections lead to superior segmentation performance. Furthermore, an adaptive depth selection mechanism is applied, which eliminates the need to train multiple models of different depths, thereby improving computational efficiency.

In order to improve the image inpainting effects of critical areas, a modified mean squared error loss (nn.MSELoss) with a weighting function was used. The formula of the weight function is: $W = \alpha \cdot M +1$, where $M$ is the CR binary mask and $\alpha$ is a tunable coefficient. By setting different weights in different regions, it achieves optimal balance through two components: a local correction term( $\alpha \cdot M$), that focuses more on contaiminated regions ( with $\alpha = 4.0$, optimized via grid search over $\alpha \in [1.0,5.0] $), and a global baseline term that preserves background structure and avoids local overfitting.

Our approach consists of the following two components in detail:

\begin{itemize}
\item[-]
Inpaint of the target source region: As the CR rejection model is exclusively trained on star/galaxy regions, the inpainting pipeline begins with source detection (via the CSST pipeline). For each detected source, a 64×64-pixel region centered on the source is cropped. If any pixels in the region are masked as CRs, the machine learning model will be used to inpaint those flagged pixels.
\item[-]
Inpaint of the background region: CRs in the background regions are inpainted by median filtering. The kernel size is dynamically adjusted from smaller to larger dimensions (3×3 to 11×11 pixels) based on the CR contamination area. This iterative process applies these multiple median filtering passes to inpaint CRs across different scales, with all operations strictly limited to CR-masked regions similarly.

\end{itemize}

It is worth noting that CR inpainting is performed only on the image data before photometry; this step is designed to produce more accurate photometric measurements and does not alter the Level 1 image data.

\section{Result}

\subsection{CR Detection}\label{sec:detection-result}

\begin{figure*}[ht!]
\plotone{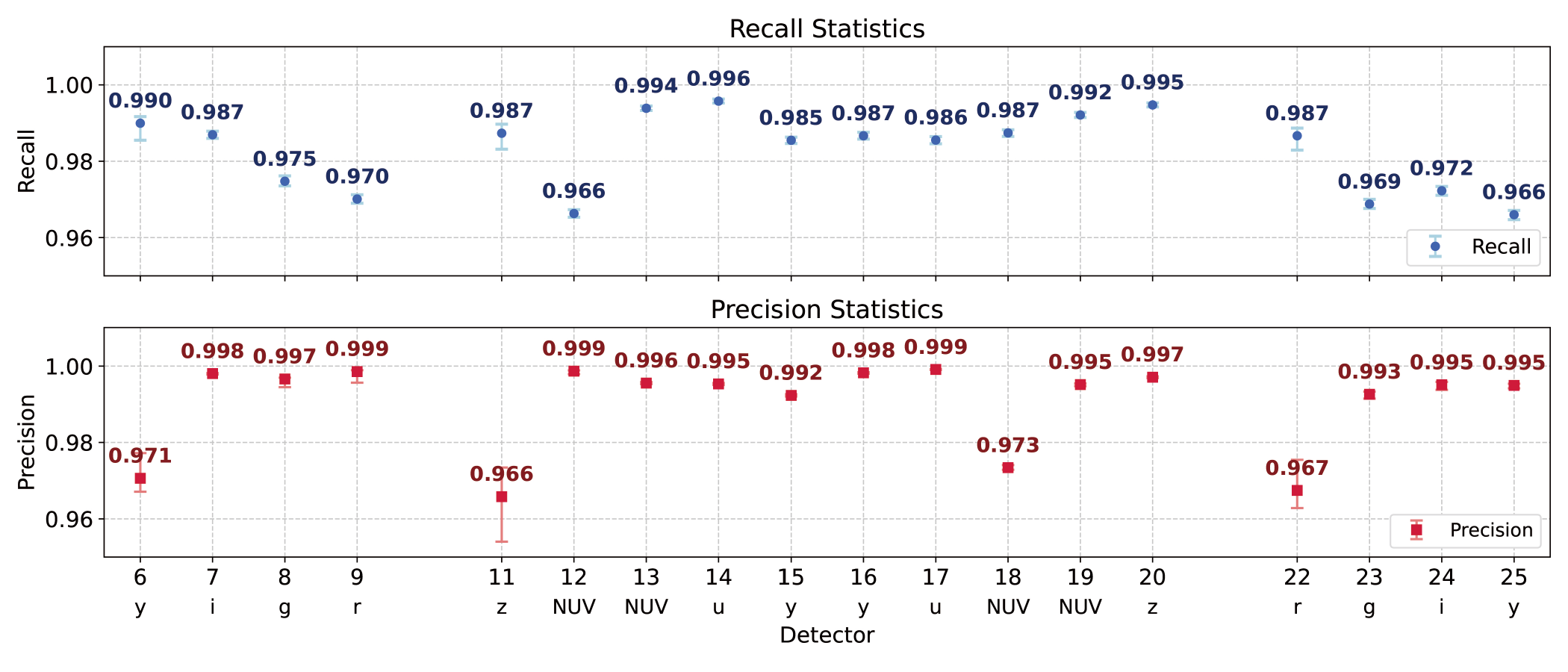}
\caption{Recall (top) and precision (bottom) of CR detection across 18 detectors. Data points and numbers represent median values calculated from 1,221 experimental trials, with asymmetric error bars indicating the 25th–75th percentile range. Detector indices (6–25) are shown on the x-axis, with missing detectors 10 and 21 identified as slitless spectroscopic instruments.}
\label{fig:detection}
\end{figure*}

We systematically analyzed CR detection performance across 21,978 CSST simulated images. During model training, we exclusively selected CRs with intensities \textgreater 1 $e^-$ as identification targets. For evaluation, we implemented more lenient criteria to assess detection performance comprehensively. Considering the model's potential sensitivity to faint 1 $e^-$ CRs, for precision metrics, we included all CRs with intensities \textgreater 0 $e^-$ to calculate false positives. Meanwhile, given the difficulty in distinguishing low-intensity CRs from background noise, recall rates were calculated using only CRs with intensities \textgreater 5 $e^-$ (corresponding to the readout noise (RON) level) to mitigate noise contamination. Figure \ref{fig:detection} summarizes the recall and precision distributions across all 18 detectors. The recall rate indicates the proportion of successfully detected CRs, demonstrating the capability of CR detection. And the (1 - precision) indicates how many pixels without CR contamination are were incorrectly identified as CRs.

The CR detection system demonstrates outstanding overall performance on CSST simulated datasets. The retrained DeepCR model achieves a recall of 97.90 ± 0.18\% (mean±SD) and precision of 98.67 ± 0.05\%, which closely matches DeepCR's performance on HST data. This result confirms that DeepCR is effective for CSST applications. On different images, the difference in detection performance of CRs comes from two aspects: hardware differences (e.g., quantum efficiency) between detectors and observational conditions (e.g., stellar density, noise) within individual detectors. Upon inspecting those low-precision cases, we found that the false positives mainly come from other noise sources, such as streak noise and oversaturated pixels in images. Notably, machine learning can recognize the PSF distribution characteristics of real sources, which is an important feature that distinguishes stars from CRs; thus, there is no concern that small-scale stars might be misclassified as CRs.

\begin{figure*}[ht!]
\centering
\includegraphics[width=0.99\textwidth]{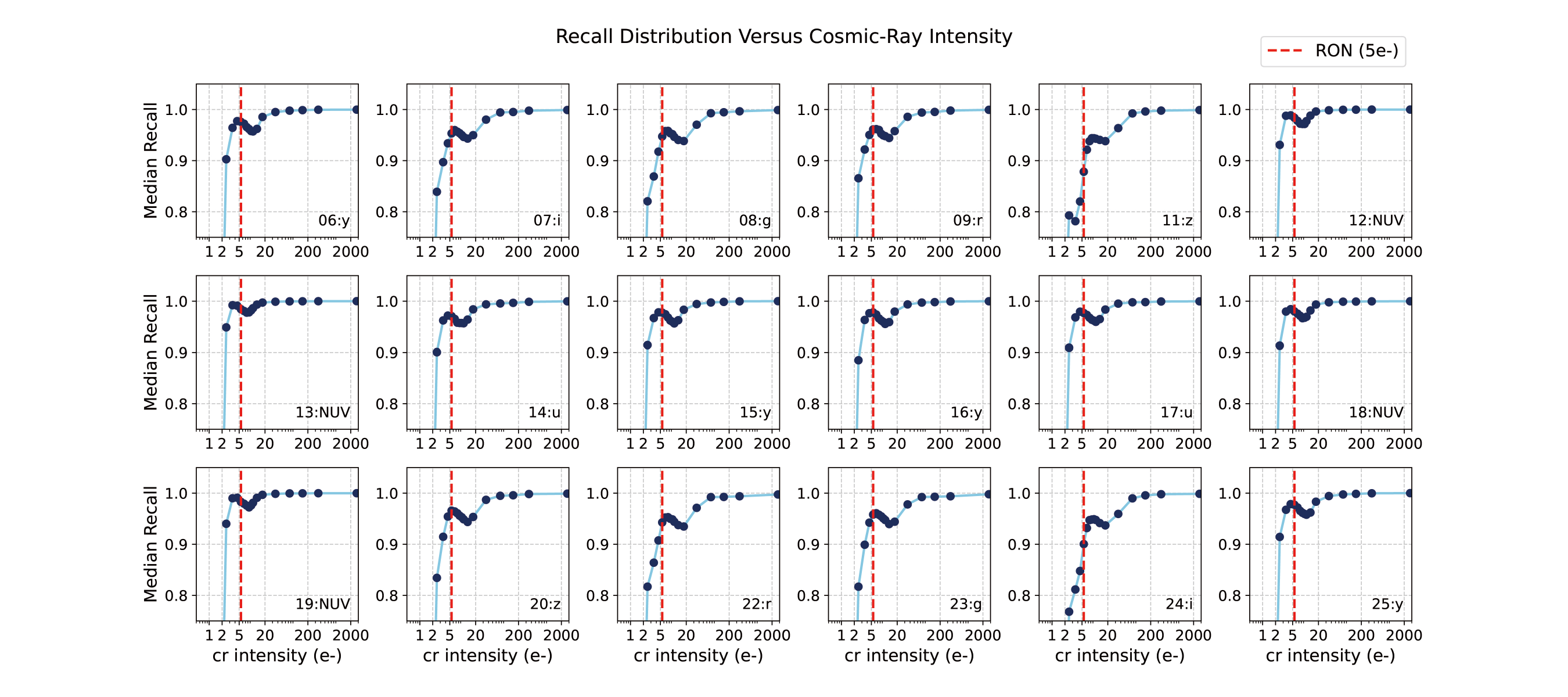}
\caption{Recall distribution versus CR intensity. Each panel corresponds to one detector, showing median recall rates from 1,221 exposure images. The horizontal axis displays CR intensity on a logarithmic scale. The vertical axis shows recall rate (uniform range: 0.75-1.0) for cross-detector comparison. The red dashed line marks the $5 e^{-}$ readout noise (RON) threshold.}
\label{fig:recall}
\end{figure*}

To further investigate CR detection performance, we analyzed recall rates across varying CR intensities (in unit of $e^-$ pixel$^{-1}$). As shown in Figure \ref{fig:recall}, recall tends to 1 for all detectors as CR intensity increases, indicating reliable identification of significantly affected pixels.

\begin{figure*}[ht!]
\plotone{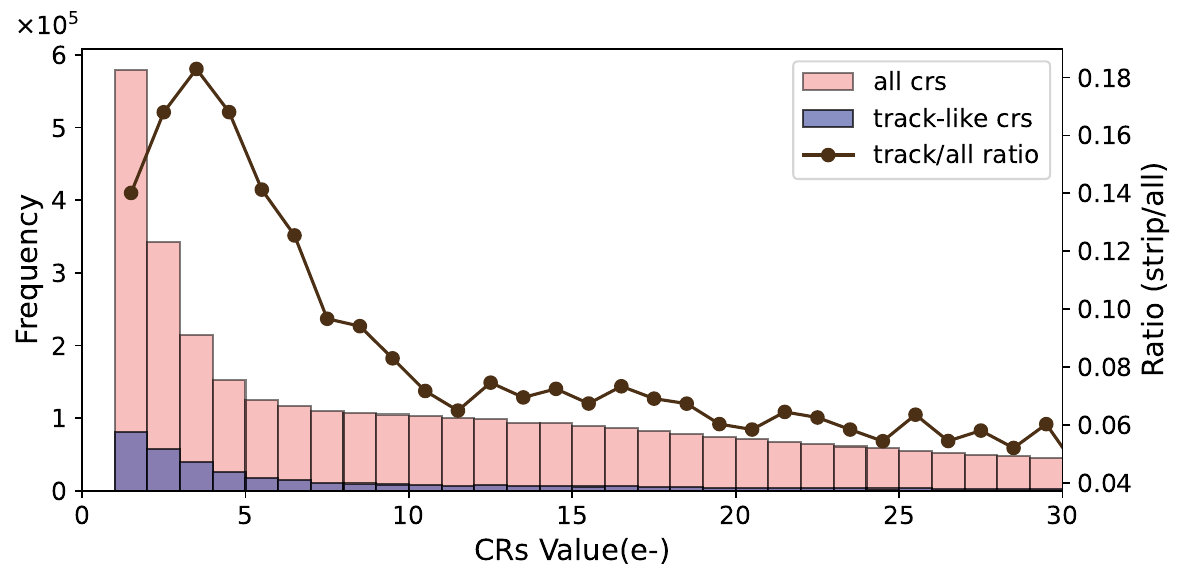}
\caption{CR intensity distribution from simulated images. The left y-axis displays frequency distributions: brown bars represent all CRs, while purple bars represent track-like CRs. The right y-axis shows the ratio of track-like to all CR counts (black line).}
\label{fig:crs}
\end{figure*}

Notably, a minor recall dip occurs during this progression, indicating abnormally low recall for CRs with deposited charges of about 6-20 electrons. Therefore, we examined all the false negatives in the images and found that the false negatives primarily originates from track-like cosmic rays and the edges of point-like cosmic rays. We speculate this dip could be connected to the energy deposition distribution of CRs. Due to different incident directions, CRs will form different shapes in the image. When CRs of the same intensity hit an image, the track-like CRs' distribution range is wider, therefore their energy deposition per pixel will be lower. This makes the energy deposition distribution of track-like CRs more concentrated in low-energy areas compared with point-like CRs. Figure \ref{fig:crs} shows the statistical CR distribution of 20 images of detector 09 as an example, revealing a rapid decline in the ratio of track-like CRs, exactly at the recall dip region. DeepCR's machine learning-based detection inherently depends on training data distribution. Although higher-intensity CRs are easier to detect, this sharp drop in sample density leads to marginally incomplete training, slightly lowering recall. Consequently, we excluded 1 $e^-$ CRs (constituting 10\% of all CRs) during model training to maintain sample balance.

The experimental results conclusively demonstrate DeepCR's excellent performance in CR identification tasks. The current simulation-trained model already exhibits robust detection capability. While the current training is based on simulated images, once acquiring observational data, we will build a sufficient dedicated training set for each detector and retrain the model to address the current issues.

\subsection{CR Rejection for Individual Sources}

\begin{figure*}[ht!]
\plotone{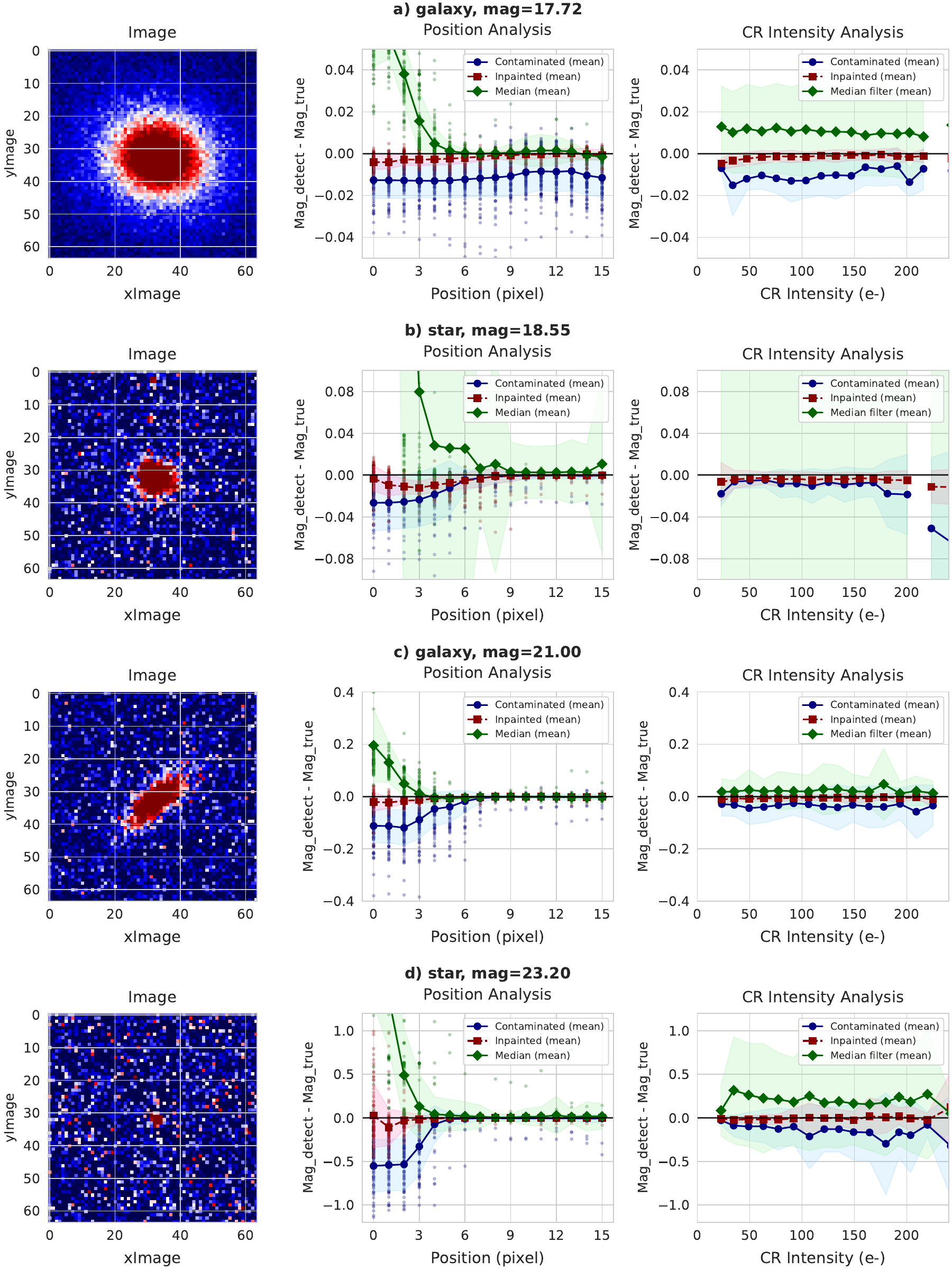}
\caption{Photometric accuracy between CR inpainting methods for individual sources. Left: original images of sources. Middle: magnitude differences versus CR distances (pixels) for CR-contaminated images (blue), corrected images (red) and median-filter corrected images (green). Solid lines show binned means, shaded regions indicate ±1 standard deviation, and raw data points display individual measurements. Right: magnitude differences versus CR intensity for the same three cases. Vertical scales adapt to magnitude ranges. 
\label{fig:inpaint_single}}
\end{figure*}

The performance of the CR rejection model on individual targets was evaluated using the simulation dataset described in Section \ref{sec:testset}. For comparative analysis, we processed the contaminated images using two different approaches: our morphology-sensitive machine learning method and the conventional median filtering method. \textit{Source Extractor} was then applied to perform photometry before and after inpainting. The magnitude difference was calculated as $\Delta mag = mag_{\text{measured}} - mag_{\text{true}}$. Relationships were established between CR position/intensity and photometric residuals (Figure \ref{fig:inpaint_single}). We selected several representative sources to display in Figure \ref{fig:inpaint_single}: panels a) and c) show galaxies; panels b) and d) show stars. The magnitudes for panels a to d are 17.72, 18.55, 21.00, and 23.20, respectively. Note that the intensity here reflects the total energy of a cosmic ray.

Our inpainting method results in unbiased magnitude differences (the median values are close to zero), showing a dramatic improvement in photometric accuracy compared to other methods. For example, in panel d) of Figure \ref{fig:inpaint_single}, where a CR impacted the center of a star, the photometric errors were -0.55 ± 0.30 (contaminated), 1.91 ± 0.95 (median filtered), and 0.03 ± 0.37 (inpainted with our method), respectively. These results clearly demonstrate that median filtering introduces large photometric errors when rejecting CRs overlapping stars or galaxies, while our method effectively addresses this issue. Meanwhile, the correction results for representative magnitudes (17, 19, 21, and 23 mag) confirm our CR rejection method's effectiveness for both stars and galaxies across source brightness levels.

We can find that both uncorrected and median-filtering-corrected cases exhibit strong positional correlations. However, as shown in Figure \ref{fig:inpaint_single}, our approach successfully mitigates the influence of CRs at varying radial distances from source centers and across diverse intensities. Notably, all three scenarios (uncorrected, median-filtered, and our corrected method) show weak correlations between CR intensity and magnitude differences. This situation likely occurs because the impacts of CRs depend more on their shape distribution, relative position to sources, and source morphology rather than intensity alone.

\subsection{Full-Frame Result}

To systematically evaluate the full-image correction efficacy of our model, we randomly selected 50 images from each of the 18 multi-band imaging detectors, then applied the CR rejection process to them. Statistical analysis of sources within these 50×18 images revealed that approximately 2.5\% of stars and 4.1\% of galaxies were contaminated by CRs. Next, we used the CSST standard pipeline to perform Kron photometry on both the original and corrected images, and calculated photometric errors by cross-matching with input catalogs.

\begin{figure*}[ht!]
\plotone{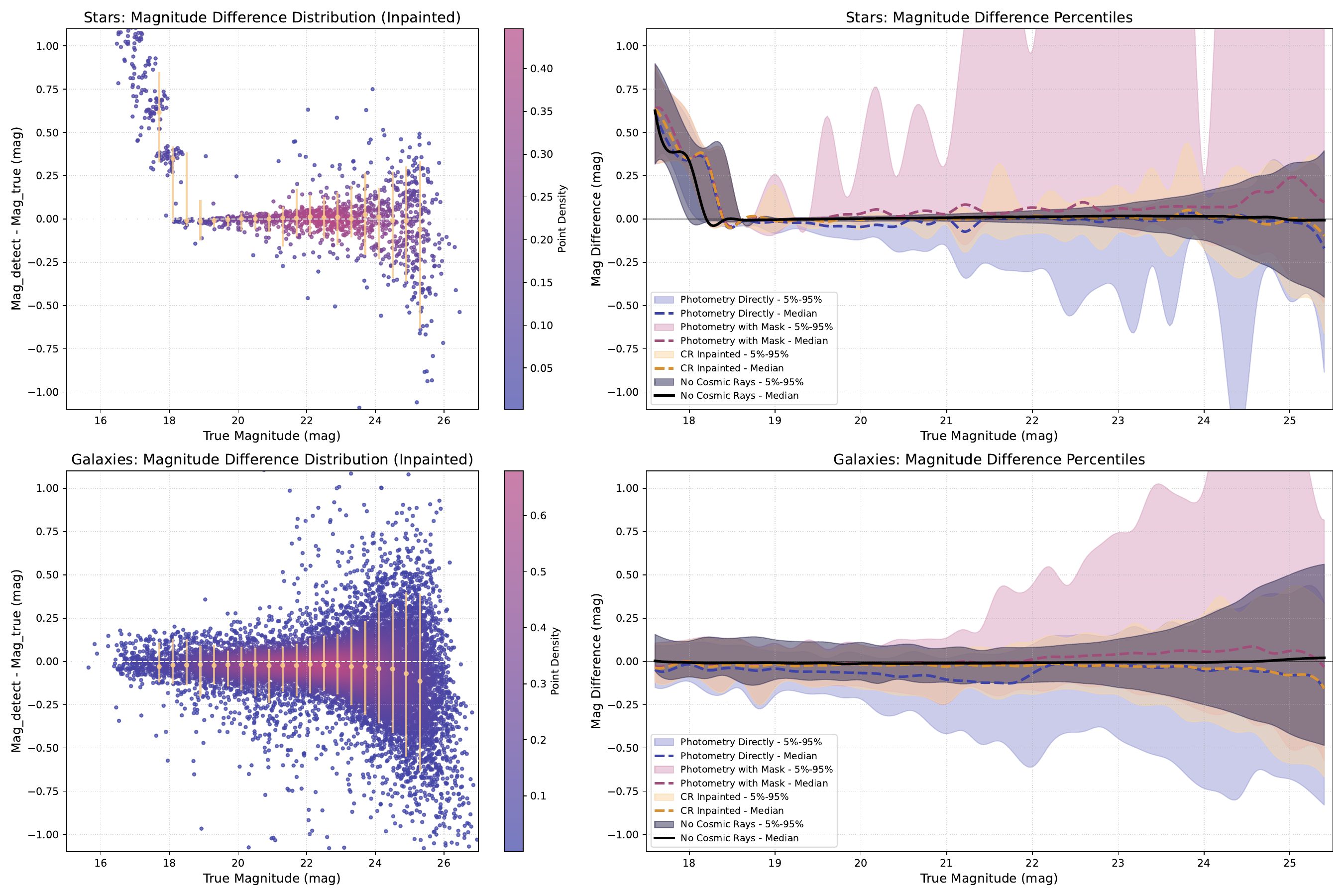}
\caption{Photometric error distributions for CR affected stars (top) and galaxies (bottom) from full frames. The left column displays scatter plots of magnitude difference (measured - true magnitude) versus true magnitude for CR-contaminated sources processed with inpainting techniques. Points are color-mapped by local density (blue: sparse, red: dense) with overlaid binned statistics (5th/50th/95th percentiles) shown as gold error bars. The right column compares percentile trends across four CR handling methods: traditional photometry without correction (blue), masked photometry (pink), inpainted photometry (orange), and CR-free reference data (black). Shaded bands represent 5th-95th percentile ranges and lines indicate medians.}
\label{fig:magdifference}
\end{figure*}

The statistical results of magnitude errors after CR rejection are plotted in the left column of Figure \ref{fig:magdifference}. Because photometric behaviors vary between galaxies and stars (such as saturation effects in stars), we subdivided the sources into stars (top panel) and galaxies (bottom panel). For each subgroup, we computed and plotted the median values along with the 5th and 95th percentiles within magnitude bins. To evaluate the effectiveness of the CR rejection, we included three groups for comparison: (1) Photometry performed while ignoring CRs (purple curve, right column). Since CRs increase flux, the magnitude result is underestimated. (2) Photometry excluding pixels that were masked as CRs (pink curve), which leads to overestimated magnitudes. (3) Uncontaminated sources (black curve), serving as a reference for the intrinsic photometric error distribution. Results after CR rejection are shown as a yellow curve. The median curve and the 5\%–95\% range are also indicated. It is evident that after CR rejection, the median values align closer to zero, and the variance is significantly reduced.

Moreover, the number of detected sources increases substantially. The number of sources identified by our machine learning approach shows a 13.6\% improvement over the masking method. These counts are derived from cross-matching measurement catalogs with the truth catalog and tabulated in Table \ref{tab:source_counts}.

\begin{table}[ht!]
\centering
\caption{Detected source counts under different CR mitigation methods}
\label{tab:source_counts}
\begin{tabular}{lccc}
\toprule
Source Type & Photometry Directly & Photometry with Mask &CR Inpainted\\
\midrule
Star        & 19,563  & 14,635  & 18,784  \\
Galaxy      & 172,979 & 166,406 & 186,907 \\
Total       & 192,921 & 181,397 & 206,064 \\
\bottomrule
\end{tabular}
\end{table}

\begin{figure*}[ht!]
\plotone{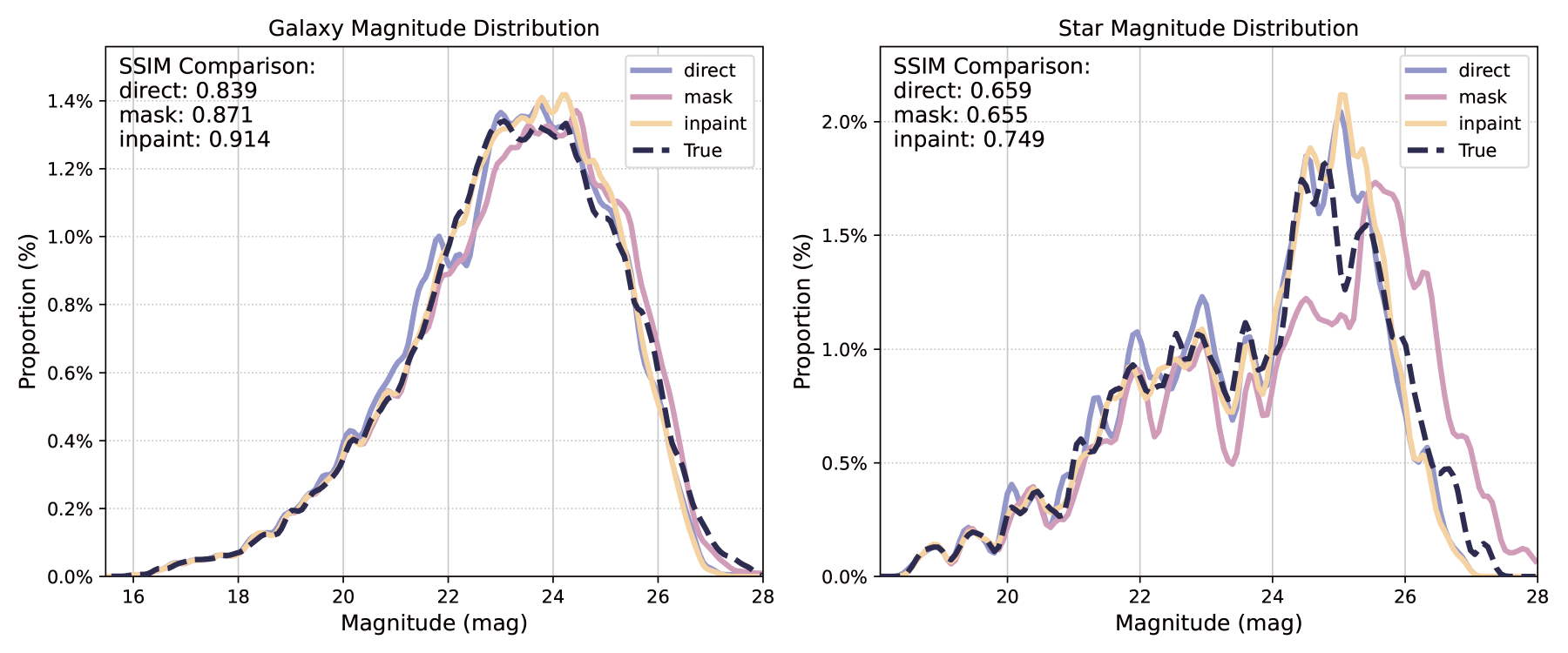}
\caption{Comparative analysis of number count performance. Magnitude distributions for stars (right) and galaxies (left) using three CR mitigation approaches: "direct" (representing uncorrected photometry ignoring CRs, light blue solid line), "mask" (denoting photometry excluding masked CR contaminated pixels, pink solid line), and "inpaint" (indicating photometry after CR inpainting, orange solid line). The dark blue dashed line represents the true magnitude distribution from the input catalog. Structural Similarity Index (SSIM) values quantifying the agreement between each mitigation method and this true distribution are displayed in the upper-left corner of each panel.}
\label{fig:counting}
\end{figure*}

Additionally, to evaluate CR rejection efficacy from a different perspective, we investigated the stellar and galactic number counts. This fundamental astronomical metric enables investigations including Galactic structure modeling, stellar population gradient analyses, and luminosity function reconstruction, while providing essential constraints for cosmological large-scale structure studies. Using Detector 08 as an illustrative example, Figure \ref{fig:counting} compares the magnitude proportion distributions of three photometry methods (direct photometry, photometry with CR mask, and photometry after CR inpaint). The ground-truth distribution (black dashed line) from the input catalog is also plotted as a reference. The Structural Similarity Index (SSIM, a perception-based metric assessing distributional resemblance where 1 indicates perfect identity) is shown in the upper left of the figure. The inpainting method achieves SSIM = 0.914 for galaxies and 0.749 for stars, demonstrating significantly closer alignment to truth than other approaches. Meanwhile, the corrected distribution exhibits near-perfect superposition with the ground truth, particularly at the bright end.

\section{Conclusion}

This paper presents the CR processing pipeline for the CSST, including CR detection, masking, and pixel-level CR rejection. For CR detection, we applied the DeepCR method, achieving a recall of 97.90±0.18\%(mean ±SD) and precision of 98.67±0.05\% on CSST simulated images. 

For CR rejection, a novel astrophysical source-targeted inpainting method is developed. We trained a dedicated Unet++ model specifically for CR-contaminated stars/galaxies, achieving accurate CR pixel rejection, and in this way substantially enhance photometric accuracy.

Compared to conventional median filtering techniques, our method demonstrates unbiased and precise photometric results for both stars and galaxies across source brightness levels, and mitigates the influence of CRs at varying radial distances from source centers and across diverse intensities.

Moreover, full-frame validation across 900 multi-band images demonstrates this method's superiority: compared to simply masking the CR contaminated pixels, our inpainting method reduces photometric errors and variance significantly; increases detectable source counts by 13.6\%; and shows near-perfect magnitude distribution agreement with ground truth (galaxy SSIM=0.914, star SSIM=0.749), particularly at bright magnitudes. 

While the current model is optimized for simulated data, we will implement transfer learning to adapt models to real observational data upon CSST commissioning. The specific method is described in detail in Section 2. By resolving the CR inpainting problem in science-critical regions, our pipeline establishes a new approach for next-generation space telescopes, showing significant potential for adaptation to other astronomical facilities facing similar CR challenges under exposure-limited conditions.

\begin{acknowledgments}
This work is based on the mock data created by the CSST Simulation Team, which is supported by the CSST scientific data processing and analysis system of the China Manned Space Project. The simulated data were reduced by the CSST scientific data processing and analysis system of the China Manned Space Project. 
This work is supported by China Manned Spaced Project (CMS-CSST-2025-A21) and the National Natural Science Foundation of China (NSFC; grants 12233008).
We are grateful to Drs. Feng Wang, Guoliang Li, and Xiaoji Lingchen for helpful discussions.
\end{acknowledgments}



%
\facilities{CSST}

\software{Source Extractor \citep{1996A&AS..117..393B}
          }






\bibliography{sample7}{}
\bibliographystyle{aasjournalv7}



\end{document}